\documentstyle[aps,aipbook,floats,psfig]{revtex}
\begin{document}

\pssilent

\title{BATSE SD Observations of \\ Hercules X-1}

\author{P.~E.~Freeman$^1$, D.~Q.~Lamb$^1$, R.~B.~Wilson$^2$, M.~S.~Briggs$^3$, 
W.~S.~Paciesas$^3$, R.~D.~Preece$^3$, D.~L.~Band$^4$, and J.~L.~Matteson$^4$}
\address{
$^1$Dept. of Astronomy and Astrophysics, University of Chicago, Chicago, IL 
60637\\
$^2$NASA Marshall Space Flight Center, Huntsville, AL 35812\\
$^3$Dept. of Physics, University of Alabama Huntsville, Huntsville, AL 35899\\
$^4$CASS, University of California San Diego, La Jolla, CA 92093\\
}

\maketitle

\begin{abstract}
The cyclotron line in the spectrum of the accretion-powered pulsar Her
X-1 offers an opportunity to assess the ability of the BATSE
Spectroscopy Detectors (SDs) to detect lines like those seen in some
GRBs.  Preliminary analysis of an initial SD pulsar mode observation
of Her X-1 indicated a cyclotron line at an energy of $\approx$ 44 keV,
rather than at the expected energy of $\approx$ 36 keV.  Our analysis
of four SD pulsar mode observations of Her X-1 made during high-states
of its 35 day cycle confirms this result.  We consider a number of
phenomenological models for the continuum spectrum and the cyclotron
line.  This ensures that we use the simplest models that adequately
describe the data, and that our results are robust.  We find modest
evidence (significance $Q \sim 10^{-4}$-$10^{-2}$) for a line at
$\approx$ 44 keV in the data of the first observation.  Joint fits to
the four observations provide stronger evidence ($Q \sim
10^{-7}$-$10^{-4}$) for the line.  Such a shift in the cyclotron line
energy of an accretion-powered pulsar is unprecedented.
\end{abstract}

\section*{Introduction}

BATSE Spectroscopy Detector (SD) observations of the cyclotron line in
the spectrum of the accretion-powered pulsar Hercules X-1 provides an
opportunity to assess the capability of the SDs to detect such lines in
the spectra of GRBs.  Observations of Her X-1 have been performed by
many groups \cite{trum78,grub82,vog82,soong90,mihara90}.  In
particular, analysis of data from the {\it HEAO 1} A4
instrument\cite{soong90}, which is a NaI detector (like the SDs) and
has an effective area and resolution that are similar to a single SD,
yields a line center energy $E$ = 36 keV and equivalent width $W_{\rm
E}$ = 8 keV.  However, preliminary analysis\cite{palm95} of an initial
SD pulsar mode observation of Her X-1 indicated a line at $\approx$ 44
keV, rather than at the expected energy of 36 keV.  We confirm this
result by applying rigorous statistical methods developed for analysis
of GRB spectra\cite{free96} to the analysis of four BATSE SD
observations of Her X-1.  A companion paper\cite{pac96} describes
studies that have been performed which confirm that the SDs are
functioning as expected.

\begin{figure}[t]
\vspace{-1.0cm}
{\leavevmode\psfig{file=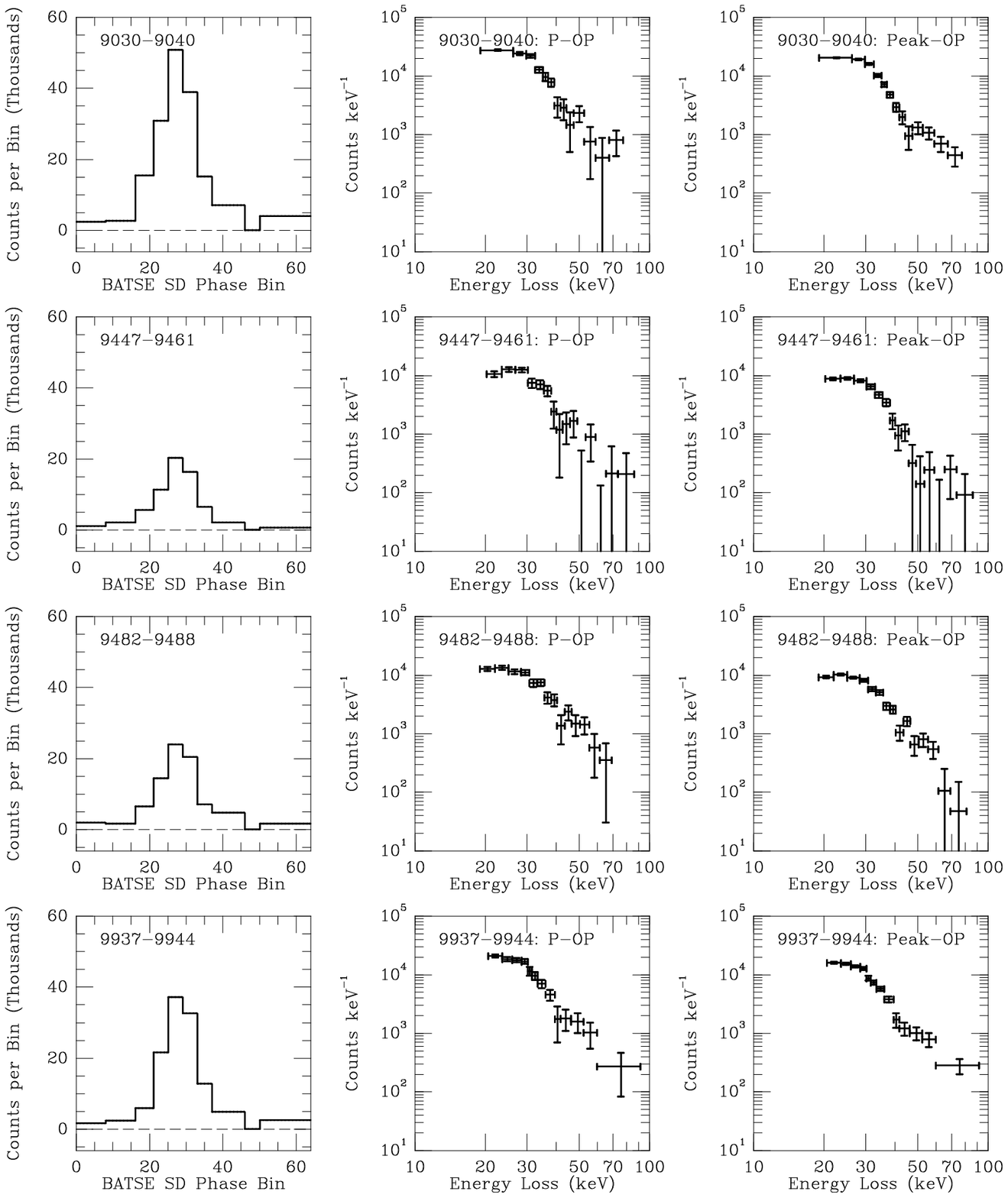,height=5.42in,width=4.55in,angle=0}}
{Figure~\ref{fig1}.
{\it Left Column:} the pulse-phase spectra for the four BATSE SD
observations.  We have re-binned the 64 SD phase bins into the 10 phase
bins defined by Soong {\it et al.}, and subtracted counts in order to
highlight the pulse.  The seven phase bins with the largest number of
counts comprise the ``On-Pulse" or P phase interval, and the remaining
bins comprise the ``Off-Pulse" or OP interval; the three phase bins
with the largest number of counts comprise the ``Peak'' interval.  {\it
Middle Column:}  the P-OP difference spectra.  {\it Right Column:}  the
Peak-OP difference spectra.
}
\refstepcounter{figure} 
{\label{fig1}}
\end{figure}

\clearpage

\begin{figure}[t]
\vspace{-0.5cm}
{\leavevmode\psfig{file=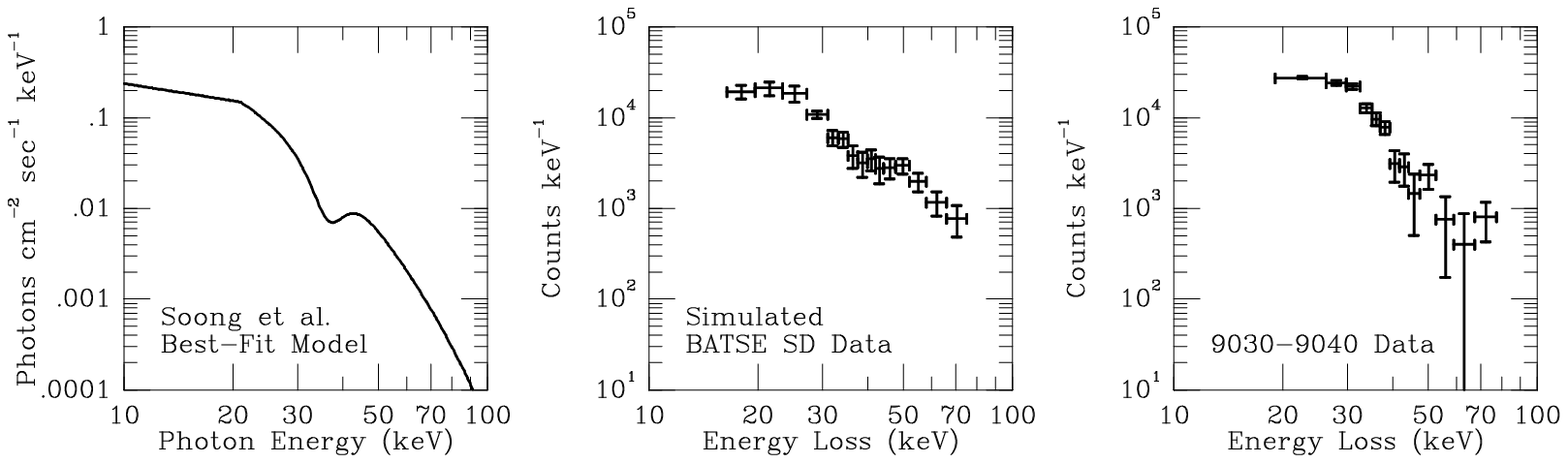,height=1.35in,width=4.55in,angle=0}}
{Figure~\ref{fig2}.  
{\it Left Panel:}  the best-fit model used by Soong {\it et al.} to fit the
P-OP difference spectrum seen by {\it HEAO 1} A4.  {\it Middle Panel:} 
the expected P-OP difference spectrum for an 80 ks observation by a
BATSE SD, created by folding the Soong {\it et al.} best-fit spectrum through
the SD response matrix.  {\it Right Panel:}  the observed P-OP 
difference spectrum for the 194 ks first observation, 9030-9040.
}
\refstepcounter{figure} 
{\label{fig2}}
\end{figure}

\begin{table}[t]
\caption{Analyzed BATSE SD Observations of Her X-1}
\label{table1}
\begin{tabular}{cccc}
Observation (TJD$^a$) & SD & $\theta_{\rm inc}$ & $t_{\rm obs}$ (ks)\\
\tableline
9030-9040 & 0 & 19$^{\circ}$ & 194 \\
9447-9461 & 6 & \phantom{1}8$^{\circ}$ & 158 \\
9482-9488 & 6 & 23$^{\circ}$ & \phantom{1}93 \\
9937-9944 & 1 & 24$^{\circ}$ & 161 \\
\end{tabular}
\noindent{$^a$ Truncated Julian Date}
{\label{tabobs}}
\vskip -7pt
\end{table}

\section*{Analysis}

\vskip -6pt
In Table~\ref{tabobs} we list the four highest quality SD observations
of Her X-1 made to date; these observations were undertaken during
high-states of the Her X-1 35 d cycle.  Figure~\ref{fig1} shows the SD
pulse-phase data rebinned into the 10 phase bins defined by Soong {\it
et al.}, and the ``On-Pulse'' minus ``Off-Pulse" (P-OP) and the``Peak''
minus ``Off-Pulse'' (Peak-OP) difference spectra.  We analyze the P-OP
difference spectra and, to improve S/N, the Peak-OP difference
spectra.  During the observations, the Low Level Discriminator (LLD)
was set to $\approx$ 10 keV.  Because of non-linearities in the
energy-to-channel conversion within $\approx$ 10 keV of the energy of
the LLD\cite{band92}, we fit only to energy-loss bins above 20 keV.

Soong {\it et al.} fit the spectrum of Her X-1 using a continuum model
which is a power law up to a break energy, and an exponential above
this energy times a Gaussian line (Figure \ref{fig2}).  The count
spectrum seen by BATSE differs in three ways from that expected from
folding the best-fit Soong et al. spectrum through the SD response
matrix and adding simulated noise: the break energy of the observed
spectrum is $\approx$ 10 keV higher than expected, the slope of the
observed spectrum above the break is greater than expected, and the
observed spectrum does not show the expected plateau around $\approx$
36 keV (Figure~\ref{fig2}).

Because the observed spectra differ from that expected, we investigate
not only the Soong {\it et al.} continuum model, but other continuum
models as well.  We seek the simplest model that adequately fits the
data.  We use a statistical criterion based on the maximum likelihood
ratio (MLR) test\footnote{ This test assumes that the difference
${\Delta}\chi^2$ resulting from two model fits to data is distributed
like $\chi^2$ for $N$ degrees of freedom, where $N$ is the number of
{\it additional} parameters in the more complicated of the two models.
$Q$ is the area under this distribution for $\chi^2 > {\Delta}\chi^2$,
and small $Q$ values indicate that it is unlikely that the improvement
${\Delta}\chi^2$ between the two fits would occur by
chance.}\cite{free96,eadie71}.  Use of this criterion often leads to
the selection of a broken power law (BPL) model for the continuum,
rather than the Soong {\it et al.} model.  Joint fits indicate that
this preference becomes stronger as we include the data from additional
observations.  The preference for the BPL over the Soong {\it et al.}
model may be a consequence of the fact that we are unable to include
data below 20 keV.  Because the line lies on the steeply-falling part
of the Her X-1 spectrum, there can be a large difference between the
values of the Soong {\it et al.} and BPL continuum models at the line,
and therefore in the significance of the line.  We therefore give the
results of continuum-plus-line fits to the data using both the Soong
{\it et al.} and BPL continuum models.

We use an exponentiated Gaussian absorption model to fit the line.  In
this model, the line full-width at half-maximum $W_{{1}\over{2}}$ =
$\eta~W_E$, and $\eta\approx$ 1 represents a saturated line.  Using a
procedure analogous to the one that we use to select continuum models,
we select the simplest continuum-plus-line model that adequately fits
the data.  We find that a one-parameter saturated line model in which
we fix the line center energy at $\approx$ 36 keV {\it never} leads to
a reduction in $\chi^2$ from the best continuum fit.  We find that a
two-parameter saturated line model adequately fits all of the data.  We
use the MLR test to determine the significance of the line.

\begin{table}[t]
\caption{Analysis of Single Observations: Peak Phase Interval}
\label{table2}
\begin{tabular}{ccccccc}
 & \multicolumn{3}{c}{Soong} & \multicolumn{3}{c}{BPL} \\
Obs. (TJD)&$E$ (keV)&$W_{\rm E}$ (keV)&$Q$&$E$ (keV)&$W_{\rm E}$ (keV)&$Q$\\
\tableline
9030-9040&44.0&7.5&2.5$\times$10$^{-4}$&44.3&6.1&8.7$\times$10$^{-3}$ \\
9447-9461&\multicolumn{3}{c}{No Improvement in $\chi^2$}&
\multicolumn{3}{c}{No Improvement in $\chi^2$} \\
9482-9488&40.5&1.9&0.50\phantom{0}&
\multicolumn{3}{c}{No Improvement in $\chi^2$} \\
9937-9944& 42.8 & 5.6 & 0.013 & 43.6 & 4.4 & 0.068 \\
\end{tabular}
{\label{tabsing}}
\vskip -5pt
\end{table}

In fits to the P-OP difference spectra, we find no evidence for a line
in any of the four individual SD observations.  The largest line
significance ($Q$ = 0.18) occurs for the first observation.  Fits to
simulated SD data created using the best-fit Soong {\it et al.} model
for the Her X-1 spectrum, but with the line energy shifted to $\approx$
43 keV, show that this result is not inconsistent with a line at
$\approx$ 43 keV.  Using joint fits, we find that the largest line
significance ($Q$ = 0.05) occurs when we combine data from the first
and fourth observations.

\begin{table}[t]
\caption{Analysis of Joint Observations: Peak Phase Interval}
\label{table3}
\begin{tabular}{ccccccc}
 & \multicolumn{3}{c}{Soong} & \multicolumn{3}{c}{BPL} \\
Obs. (TJD)&$E$ (keV)&$W_{\rm E}$ (keV)&$Q$&$E$ (keV)&$W_{\rm E}$ (keV)&$Q$\\
\tableline
9030 & 44.0 & 7.5 & 2.5$\times$10$^{-4}$ & 44.3 & 6.1 & 8.7$\times$10$^{-3}$ \\
9030/9937&43.6&6.7&4.7$\times$10$^{-7}$&44.1&5.4&9.8$\times$10$^{-5}$ \\
All - 9482&43.8&6.7&3.2$\times$10$^{-7}$ & 44.1 & 5.3 & 8.8$\times$10$^{-5}$ \\
All & 43.1& 5.3 & 2.3$\times$10$^{-6}$ & 43.5 & 3.9 & 4.6$\times$10$^{-4}$ \\
9447/9937 & 43.7 & 5.6 & 0.010 & 44.9 & 4.8 & 0.054 \\
All - 9030&41.1 & 2.1 & 0.21 & \multicolumn{3}{c}{No Improvement in $\chi^2$}\\
\end{tabular}
{\label{tabj}}
\end{table}

In fits to the Peak-OP difference spectra, we detect a line at
$\approx$ 44 keV with modest significance in the data from the first
observation (Table~\ref{tabsing}), but not in the data from any other
observation.  Joint fits indicate that this line becomes more
significant as we add the data from more observations
(Table~\ref{tabj}).  However, the addition of data from the third
observation reduces the significance of the line.  This behavior may
reflect the fact that each of the four observations covers a slightly
different phase of the Her X-1 35 d cycle.  If we exclude the first
observation from the joint fits, we find that we cannot detect the
line.

We estimate the statistical error for each model parameter using Monte
Carlo simulations of the best-fit models.  The typical 1$\sigma$
uncertainties for both $E$ and $W_E$ are $\approx$ 1.0-1.5 keV.

\section*{Conclusions}

Joint fits to four SD observations of Her X-1 show strong evidence for
a cyclotron scattering line at $\approx$ 44 keV, rather than at the
expected energy of 36 keV.  A cyclotron line energy shift is
unprecedented in observations of accretion-powered pulsars.  A
companion paper \cite{pac96} describes studies that have been conducted
which confirm that the SDs are functioning as expected.


\begin{references}
\bibitem{trum78}J. Tr{\"u}mper, W. Pietsch, C. Reppin, W. Voges, R.
Staubert, and E. Kendizorra, Ap. J. {\bf 219}, L105 (1978).
\bibitem{grub82}D. E. Gruber, et al., Ap. J. {\bf 240}, L127 (1982).
\bibitem{vog82}W. Voges, W. Pietsch, C. Reppin, J. Tr{\"u}mper, E.
Kendizorra, and R. Staubert, Ap. J. {\bf 263}, 803 (1982).
\bibitem{soong90}Y. Soong, et al., Ap. J. {\bf 348}, 641 (1990).
\bibitem{mihara90}T. Mihara, K. Makashima, T. Ohashi, T. Sakao, and M.
Tashiro, Nature {\bf 335}, 234 (1990).
\bibitem{palm95} D. M. Palmer, et al., Gamma-Ray Bursts, eds. G. J.
Fishman, J. J. Brainerd, and K. Hurley (New York: AIP, 1994), p. 247.
\bibitem{free96}P. E. Freeman, et al., Ap. J., {\it submitted}.
\bibitem{pac96}W. S. Paciesas, et al., these proceedings.
\bibitem{band92}D. Band, et al., Exp. Astr. {\bf 2}, 307 (1992).
\bibitem{eadie71}W. T. Eadie, D. Drijard, F. E. James, M. Roos, and
B. Sadoulet, Statistical Methods in Experimental Physics (Amsterdam: 
North Holland, 1971).
\end{references}
\end{document}